\documentclass[12pt,letterpaper]{article}
\usepackage{amsmath,amssymb,array,calc,rotating,epsfig,psfrag,amscd, cite}

\makeatletter
\renewcommand\section{\@startsection {section}{1}{\z@}%
                                   {-3.5ex \@plus -1ex \@minus -.2ex}%nn
                                   {2.3ex \@plus.2ex}%
                                   {\normalfont\large\bfseries}}
\renewcommand\subsection{\@startsection{subsection}{2}{\z@}%
                                     {-3.25ex\@plus -1ex \@minus -.2ex}%
                                     {1.5ex \@plus .2ex}%
                                     {\normalfont\bfseries}}
\makeatother

\let\non\nonumber

\let\s=\sigma

\let\x=\xi
\let\S=\Sigma

\newcommand{\bea}{\begin{eqnarray}}
\newcommand{\eea}{\end{eqnarray}}
\newcommand{\be}{\begin{equation}}
\newcommand{\ee}{\end{equation}}

\newcommand{\p}{\partial}

\newcommand{\C}[1]{$(\ref{#1})$}

\def\IZ{\relax\ifmmode\mathchoice
{\hbox{\cmss Z\kern-.4em Z}}{\hbox{\cmss Z\kern-.4em Z}}
{\lower.9pt\hbox{\cmsss Z\kern-.4em Z}} {\lower1.2pt\hbox{\cmsss
Z\kern-.4em Z}}\else{\cmss Z\kern-.4em Z}\fi}
\def\IR{\relax{\rm I\kern-.18em R}}

\def\one{{\hbox{ 1\kern-.8mm l}}}

\newlength{\bredde}
\def\slash#1{\settowidth{\bredde}{$#1$}\ifmmode\,\raisebox{.15ex}{/}
\hspace*{-\bredde} #1\else$\,\raisebox{.15ex}{/}\hspace*{-\bredde}
#1$\fi}

\newsavebox{\zzzbar}
\sbox{\zzzbar}
  {\setlength{\unitlength}{0.9em}
  \begin{picture}(0.6,0.7)
  \thinlines
  \put(0,0){\line(1,0){0.6}}
  \put(0,0.75){\line(1,0){0.575}}
  \multiput(0,0)(0.0125,0.025){30}{\rule{0.3pt}{0.3pt}}
  \multiput(0.2,0)(0.0125,0.025){30}{\rule{0.3pt}{0.3pt}}
  \put(0,0.75){\line(0,-1){0.15}}
  \put(0.015,0.75){\line(0,-1){0.1}}
  \put(0.03,0.75){\line(0,-1){0.075}}
  \put(0.045,0.75){\line(0,-1){0.05}}
  \put(0.05,0.75){\line(0,-1){0.025}}
  \put(0.6,0){\line(0,1){0.15}}
  \put(0.585,0){\line(0,1){0.1}}
  \put(0.57,0){\line(0,1){0.075}}
  \put(0.555,0){\line(0,1){0.05}}
  \put(0.55,0){\line(0,1){0.025}}
  \end{picture}}

\newcommand{\ena}{\end{eqnarray}}
\newcommand{\beqa}{\begin{eqnarray}}
\newcommand{\eeqa}{\end{eqnarray}}

\newcommand{\g}{\gamma}

\def\g{\gamma}

\def\s{\sigma}

\def\x{\xi}

\def\S{\Sigma}

\begin{document}
\begin{titlepage}

\begin{center}

%\hfill \today
%\hfill         \phantom{xxx}         

%\hfill HRI

\vskip 2 cm
{\Large \bf Non--BPS interactions from the type II one loop four graviton amplitude}\\
\vskip 1.25 cm { Anirban Basu\footnote{email address:
    anirbanbasu@hri.res.in} } \\
{\vskip 0.5cm Harish--Chandra Research Institute, Chhatnag Road, Jhusi,\\
Allahabad 211019, India\\}

\end{center}

\vskip 2 cm

\begin{abstract}
\baselineskip=18pt

We obtain T--duality invariant second order differential equations satisfied by the $D^8\mathcal{R}^4$ and $D^{10} \mathcal{R}^4$ interactions from the low energy expansion of the one loop four graviton amplitude in toroidally compactified type II string theory. The eigenvalues of these equations are completely determined by the structure of the one loop integrands. Unlike the BPS interactions, these non--BPS interactions satisfy Poisson equations having source terms that receive contributions from both the bulk and boundary of the worldsheet moduli space. We explicitly solve these equations in nine dimensions.

\end{abstract}

\end{titlepage}

%\pagestyle{plain}
%\baselineskip=18pt
% Try a wider skip
%\baselineskip=19pt
%%%%%%%%%%%%%%%%%%%%%%%%%%%%%%%%%%%%%%%%%%%%%%%%%%%%%%%%%%%%%%%%%%%%%%%%%%%%%%

\section{Introduction}

Maximally supersymmetric string theories provide a very useful arena for quantitatively analyzing in detail the duality symmetries of string theory. S--matrices in these theories in the Einstein frame are U--duality covariant, the determination of which yield complete information about the perturbative and non--perturbative contributions to these scattering amplitudes. They lead to moduli dependent coefficient functions for various interactions in the low energy effective action. For toroidally compactified type II string theory, several BPS interactions can be obtained exactly using both worldsheet and spacetime techniques~\cite{Green:1997tv,Green:1997as,Kiritsis:1997em,Green:1998by,Green:1999pu,D'Hoker:2005jc,D'Hoker:2005ht,Berkovits:2005ng,Green:2005ba,Berkovits:2006vc,Basu:2007ru,Basu:2007ck,Green:2008bf,Basu:2008cf,Green:2010kv,Green:2010wi,Basu:2011he,D'Hoker:2013eea,Gomez:2013sla,D'Hoker:2014gfa,Basu:2014hsa,Bossard:2014aea,Pioline:2015yea,Bossard:2015uga,Bossard:2015oxa,Basu:2015dqa,Pioline:2015nfa,Bossard:2015foa}. This is possible essentially due to the large amount of supersymmetry the theory possesses. These BPS interactions receive perturbative contributions only upto a finite number of loops and hence satisfy various non--renormalization theorems. However, the structure of the non--BPS interactions is considerably more complicated as they are expected to receive contributions from all orders in superstring perturbation theory, and hence are not expected to be given by coefficient functions which satisfy simple equations. Hence understanding the various non--BPS interactions in the effective action is important from the point of view of understanding the structure of generic interactions which are not protected by supersymmetry. 

We shall be interested in the one loop perturbative contribution to some of the simplest local non--BPS interactions that are obtained from the low energy expansion of string amplitudes. These are obtained from the low momentum expansion of the four graviton amplitude in type II string theory toroidally compactified on $T^d$, where we take $d \leq 7$. While the first three terms in the expansion give the $\mathcal{R}^4$, $D^4\mathcal{R}^4$ and $D^6\mathcal{R}^4$ interactions which are $1/2$, $1/4$ and $1/8$ BPS respectively, the remaining terms of the form $D^{2k} \mathcal{R}^4$ for $k \geq 4$ are non--BPS. We shall consider the $D^8\mathcal{R}^4$ and $D^{10}\mathcal{R}^4$ interactions which are the leading non--BPS interactions in the low momentum expansion, though the primary logic of the analysis can be extended to all non--BPS interactions. At tree level, all these interactions are simply obtained by expanding the four graviton amplitude leading to coefficients given by Riemann zeta functions. At one loop, they involve an integral over the (super)moduli space of the genus one Riemann surface. 

We show that the moduli dependent coefficient functions of these interactions satisfy T--duality invariant second order differential equations with source terms\footnote{The moduli of the compactification are given by $G_{ij}$ and $B_{ij}$, which are the metric and the NS--NS two form along the directions of $T^d$.}. These $O(d,d,\mathbb{Z})$ invariant differential equations for compactification on $T^d$ have eigenvalues that are completely determined by the structure of the one loop integrands. The source terms in these Poisson equations are of two types: those that receive contributions only from the boundary of moduli space, and those that receive contributions from the bulk of moduli space as well. Those of the first kind are easy to evaluate as they involve only a knowledge of the asymptotic nature of the various integrands, in fact for the BPS interactions the source terms are only of this type~\cite{Basu:2015dqa}. On the other hand, the source terms of the second kind are considerably more involved to obtain, which signifies the complications that arise for interactions that are not protected by supersymmetry. This happens because of the involved nature of the integrands for the various non--BPS interactions compared to the BPS ones. Note that directly obtaining the coefficient functions for these interactions by integrating over moduli space is considerably involved because of the complicated nature of the various integrands. However, obtaining differential equations is simpler as they involve only the knowledge of worldsheet modular invariant differential equations satisfied by the most complicated terms in the integrand. Thus the main objective is along the lines of~\cite{Basu:2015dqa} with the aim of looking at some simple non--BPS interactions. 
    
We start with a brief review of the one loop four graviton amplitude in toroidally compactified type II string theory mentioning facts relevant for our purposes. We then obtain the Poisson equation satisfied by the $D^8\mathcal{R}^4$ interaction, and solve it explicitly for the simplest case of compactification on a circle. We next perform a similar analysis for the $D^{10} \mathcal{R}^4$ interaction. We also briefly discuss possible non--analytic U--duality invariant terms in the effective action based on the structure of harmonic anomalies in the Poisson equations. Though we have looked at some simple examples to illustrate the main idea of solving for the various interactions by obtaining differential equations satisfied by them, our methods are quite general, and should be useful to analyze interactions at higher orders in the low momentum expansion at one loop, and also at higher loops. It should also be useful to analyze interactions in theories in other dimensions, and with less supersymmetry.

\section{The one loop four graviton amplitude}

The one loop four graviton amplitude in type II superstring theory toroidally compactified on $T^d$ for $d \leq 7$ is given by~\cite{Green:1981yb}
\be \mathcal{A}_4 = 2\pi \mathcal{I}(s,t,u) \mathcal{R}^4, \ee
where
\be \label{oneloop}\mathcal{I} (s,t,u) = \int_{\mathcal{F}} \frac{d^2\Omega}{\Omega_2^2} F(s,t,u;\Omega,\bar\Omega)\Gamma_{d,d;1},\ee
where we have integrated over $\mathcal{F}$, the fundamental domain of $SL(2,\mathbb{Z})$. Here $\Omega$ is the complex structure of the worldsheet torus. The Mandelstam variables for the external momenta $s, t, u$ satisfy the on--shell condition
\be s+t+u=0.\ee 
We have defined the measure $d^2 \Omega = d\Omega_1 d\Omega_2$. The lattice factor in \C{oneloop} denoted by $\Gamma_{d,d;1}$ at genus one for compactification on $T^d$ is given by
\be \label{genlat}\Gamma_{d,d;1} (G,B;\Omega)= V_d \sum_{m^i,n^i\in \mathbb{Z}} e^{-\pi(G+B)_{ij}(m^i +\Omega n^i)(m^j +\bar\Omega n^j)/{\Omega_2}},\ee
where $i,j=1,\ldots,d$ and $V_d$ is the dimensionless volume (in units of $l_s$) of $T^d$ in the string frame. Here $G_{ij}$ and $B_{ij}$ are the components of the metric and the NS--NS two form along $T^d$ respectively. Apart from the dependence on the worldsheet moduli, the entire spacetime moduli dependence is contained in the lattice factor. 

The momentum dependent factor $F(s,t,u;\Omega,\bar\Omega)$ in \C{oneloop} which also encodes the worldsheet moduli dependence is given by
\be \label{D}F(s,t,u;\Omega,\bar\Omega) = \prod_{i=1}^4 \int_\S  \frac{d^2 z^{(i)}}{\Omega_2} e^{\mathcal{D}}.\ee
Here $z^{(i)}$ $(i=1,2,3,4)$ are the positions of insertions of the four graviton vertex operators on the toroidal worldsheet $\S$. Hence $d^2 z^{(i)} = d({\rm Re} z^{(i)}) d({\rm Im}z^{(i)})$, where
\be -\frac{1}{2} \leq {\rm Re} z^{(i)} \leq \frac{1}{2}, \quad 0 \leq {\rm Im} z^{(i)}\leq \Omega_2 \ee
for all $i$. In \C{D}, the expression for $\mathcal{D}$ is given by
\be \label{defD}4\mathcal{D} = \alpha' s (\hat{G}_{12} + \hat{G}_{34})+\alpha' t (\hat{G}_{14} + \hat{G}_{23})+ \alpha' u (\hat{G}_{13} +\hat{G}_{24}),\ee
where $\hat{G}_{ij}$ is the scalar Green function on the torus with complex structure $\Omega$ between points $z^{(i)}$ and $z^{(j)}$, and so
\be \hat{G}_{ij} \equiv \hat{G}(z^{(i)} - z^{(j)};\Omega).\ee
In particular, it is defined as~\cite{Green:1999pv}
\bea \label{prop}\hat{G}(z;\Omega) &=& -{\rm ln} \Big\vert \frac{\theta_1 (z\vert\Omega)}{\theta_1'(0\vert\Omega)} \Big\vert^2 + \frac{2\pi({\rm Im}z)^2}{\Omega_2} \non \\ &=& \frac{1}{\pi} \sum_{(m,n)\neq(0,0)} \frac{\Omega_2}{\vert m\Omega+n\vert^2} e^{\pi[\bar{z}(m\Omega+n)-z(m\bar\Omega+n)]/\Omega_2} + 2{\rm ln} \vert \sqrt{2\pi} \eta(\Omega)\vert^2.\eea
Now the $z$ independent zero mode part given by the second term in the second line of \C{prop} cancels in the whole amplitude, which follows from the expression for $\mathcal{D}$ in \C{defD} on using $s+t+u=0$. Thus in the expression for $\mathcal{D}$ we simply replace $\hat{G}(z;\Omega)$ by $G(z;\Omega)$ where
\be \label{Green}G(z;\Omega) = \frac{1}{\pi} \sum_{(m,n)\neq(0,0)} \frac{\Omega_2}{\vert m\Omega+n\vert^2} e^{\pi[\bar{z}(m\Omega+n)-z(m\bar\Omega+n)]/\Omega_2}.\ee
To obtain the parts of the amplitude which are analytic and non--analytic in the external momenta, in \C{oneloop}, $\mathcal{F}$ is split into
\be \mathcal{F} = \mathcal{F}_L +\mathcal{R}_L,\ee
where $\mathcal{F}_L$ is defined for $\tau_2 \leq L$, and $\mathcal{R}_L$ is defined for $\tau_2 > L$~\cite{Green:1999pv,Green:2008uj}, and we take $L\rightarrow \infty$. Thus the analytic part of the amplitude is obtained by keeping the finite terms as $L\rightarrow \infty$ in $\mathcal{F}_L$, while the non--analytic part is obtained by keeping finite terms in $\mathcal{R}_L$. While both contributions can contain terms that diverge as $L\rightarrow \infty$, they cancel in the final answer.  

Thus performing a low momentum expansion, the analytic part of the amplitude is given by
\be \mathcal{I}_{an} (s,t,u) =   \sum_{n=0}^\infty\int_{\mathcal{F}_L} \frac{d^2\Omega}{\Omega_2^2}\prod_{i=1}^4 \int_\S  \frac{d^2 z^{(i)}}{\Omega_2} \cdot \frac{\mathcal{D}^n}{n!}\Gamma_{d,d;1}.\ee
Hence performing an $\alpha'$ expansion we get that
\be \mathcal{I}_{an} (s,t,u)= \sum_{p,q} \s_2^p \s_3^q J^{p,q},\ee
where
\be J^{p,q} = \int_{\mathcal{F}_L} \frac{d^2\Omega}{\Omega_2^2} j^{p,q} (\Omega,\bar\Omega)\Gamma_{d,d;1},\ee
and
\be \s_2 = \alpha'^2 (s^2 +t^2 +u^2), \quad \s_3 = \alpha'^3 (s^3 +t^3 +u^3).\ee
Here $j^{p,q}(\Omega,\bar\Omega)$ is obtained after integrating over the insertion points of the vertex operators and encodes the topologically distinct ways the scalar propagators are connected on the toroidal worldsheet.

To obtain expressions for $j^{(p,q)}$, it is convenient to
define the non--holomorphic Eisenstein series given by
$E_s (\Omega,\bar\Omega)$ as
\bea \label{Eisenstein}  E_s(\Omega,\bar\Omega) &&= \sum_{l_i \in \mathbb{Z}, (l_1,l_2) \neq (0,0)} \frac{\Omega_2^s}{\pi^s \vert l_1 + l_2 \Omega\vert^{2s}} \non \\ && = \frac{2}{\pi^s}\zeta(2s) \Omega_2^s + 2 \Omega_2^{1-s} \frac{\Gamma(s-1/2)}{\pi^{s-1/2}\Gamma(s)}\zeta(2s-1)\non \\ && +\frac{4 \sqrt{\Omega_2}}{\Gamma(s)} \sum_{k\in \mathbb{Z}, k\neq 0} \vert k \vert^{s-1/2} \mu(\vert k\vert, s) K_{s-1/2} (2\pi \Omega_2 \vert k \vert) e^{2\pi i k\Omega_1},\eea
where
\be \mu (k,s) = \sum_{m>0,m|k} \frac{1}{m^{2s-1}}.\ee
Now $E_s (\Omega,\bar\Omega)$ satisfies the Laplace equation
\be \label{eisenstein}\Delta_\Omega E_s (\Omega,\bar\Omega) = s(s-1) E_s (\Omega,\bar\Omega)\ee
which we shall often use.
Here the $SL(2,\mathbb{Z})$ invariant Laplacian is defined by
\be \Delta_\Omega = 4\Omega_2^2\frac{\p^2}{\p\Omega\p\bar\Omega}.\ee

For the leading non--BPS interactions, we have that~\cite{Green:2008uj}
\bea \label{list} j^{(2,0)} &=& \frac{1}{4!}(9 D_2^2 + 6 D_{1,1,1,1} + D_4), \non \\ j^{(1,1)} &=& \frac{5}{6!} (2 D_5 + 96 D_2 D_{1,1,1} + 28 D_2 D_3 + 32 D_{3,1,1} \non \\ &&-24 D_{2,2,1} + 24 D_{2,1,1,1} - 48 D_{1,1,1,1;1}).\eea

We now define the various quantities in \C{list}. For brevity, we denote
\be \int_{\S} d^2 z \int_{\S} d^2 w \cdots = \int_{zw\cdots}. \ee

We have that
\bea && D_l = \int_{12} G_{12}^l, \quad D_{1,1,1} = \int_{123} G_{12} G_{23} G_{31}, \quad D_{1,1,1,1} = \int_{1234} G_{12} G_{23} G_{34} G_{41}, \non \\ && D_{3,1,1} = \int_{123} G_{12} G_{23} G^3_{13}, \quad D_{2,2,1} = \int_{123} G_{12}^2 G_{13}^2 G_{23}, \quad D_{2,1,1,1} = \int_{1234} G_{12} G_{23} G_{34} G^2_{14}, \non \\  && D_{1,1,1,1;1} = \int_{1234} G_{12} G_{23} G_{34} G_{41} G_{13}.\eea
Hence these quantities encode the various ways the scalar propagators are connected on the worldsheet.  

\subsection{The BPS one loop four graviton interactions}

First let us consider the BPS interactions. The coefficient of the $\mathcal{R}^4$ interaction is given by\footnote{We denote the coefficient of the $D^{2k}\mathcal{R}^4$ interaction at genus $g$ by $I_{D^{2k}\mathcal{R}^4}^{(g)}.$}
\be \label{1genus1}I_{\mathcal{R}^4}^{(1)} = 2\pi  \int_{\mathcal{F}_L} \frac{d^2\Omega}{\Omega_2^2} \Gamma_{d,d;1} ,\ee
while the coefficient of the $D^4\mathcal{R}^4$ interaction is given by 
\be \label{1genus2}I_{D^4 \mathcal{R}^4}^{(1)} =2\pi\int_{\mathcal{F}_L} \frac{d^2\Omega}{\Omega_2^2}  E_2 (\Omega, \bar\Omega)\Gamma_{d,d;1}.\ee
Finally the coefficient of the $D^6\mathcal{R}^4$ interaction is given by
\be \label{1genus3} I_{D^6 \mathcal{R}^4}^{(1)} = \frac{2\pi}{3}\int_{\mathcal{F}_L}  \frac{d^2\Omega}{\Omega_2^2} \Big(5 E_3 (\Omega,\bar\Omega) + \zeta(3)\Big) \Gamma_{d,d;1}= \hat{I}_{D^6\mathcal{R}^4}^{(1)}+\frac{\zeta(3)}{3} I_{\mathcal{R}^4}^{(1)}. \ee

On using the relation~\cite{Obers:1999um}
\be \label{main}\Big(\Delta_{O(d,d,\mathbb{Z})} +\frac{d(d-2)}{2}\Big)\Gamma_{d,d;1} = 2\Delta_\Omega \Gamma_{d,d;1}\ee
where $\Delta_{O(d,d,\mathbb{Z})}$ is the $O(d,d,\mathbb{Z})$ invariant Laplacian, we have that~\cite{Basu:2015dqa}
\bea \Big(\Delta_{O(d,d,\mathbb{Z})} +\frac{d(d-2)}{2}\Big) I_{\mathcal{R}^4}^{(1)} &=&4\pi \delta_{d,2}, \non \\
\Big(\Delta_{O(d,d,\mathbb{Z})} +\frac{(d+2)(d-4)}{2}\Big) I_{D^4 \mathcal{R}^4}^{(1)}  &=& 12\zeta(3) \delta_{d,4},\non \\
\Big(\Delta_{O(d,d,\mathbb{Z})} +\frac{(d+4)(d-6)}{2}\Big) \hat{I}_{D^6 \mathcal{R}^4}^{(1)} &=& \frac{25\zeta(5)}{\pi} \delta_{d,6}.\eea
Thus the source terms involve harmonic anomalies in specific dimensions whose structure is entirely determined by the nature of the integrand at the boundary of moduli space.

Solving these equations, in nine dimensions, we have that
\bea \label{9d} I_{\mathcal{R}^4}^{(1)} &=& 4\zeta(2) \Big( r+\frac{1}{r}\Big), \non \\ I_{D^4\mathcal{R}^4}^{(1)} &=& \frac{4}{\pi^2}\zeta(3)\zeta(4) \Big( r^3+\frac{1}{r^3}\Big), \non \\  I_{D^6 \mathcal{R}^4}^{(1)} &=& \frac{4}{3} \zeta(2)\zeta(3) \Big( r+\frac{1}{r}\Big) + \frac{10}{\pi^4} \zeta(5)\zeta(6) \Big( r^5+\frac{1}{r^5}\Big),\eea
where $r$ is the radius of the circle in the string frame. The overall coefficients are obtained by matching with the decompactification limit. The terms linear in $r$ yield the ten dimensional result, while the ones that diverge as $r\rightarrow \infty$ yield threshold corrections on summing along with an infinite number of other such diverging terms. Their structure can be fixed using supergravity\footnote{These interactions in nine dimensions can be directly obtained by simply integrating over moduli space. We obtain them using differential equations to have a unified treatment along with non--BPS interactions as well.}.

\section{The one loop $D^8\mathcal{R}^4$ interaction}

We first consider the $D^8\mathcal{R}^4$ interaction.
For the various terms needed in $j^{(2,0)}$ in \C{list}, $D_2$ and $D_{1,1,1,1}$ are simple and are given by
\be \label{e1}D_2 = E_2, \quad D_{1,1,1,1} = E_4 .\ee
However, $D_4$ is not simple and is given by~\cite{D'Hoker:2015foa,D'Hoker:2015zfa}
\be \label{e2}D_4 = 24 C_{2,1,1} -18 E_4 + 3 E_2^2,\ee
where
\be C_{2,1,1} = \int_{123} G_{12} G_{23} G^2_{13}.\ee
It satisfies the Poisson equation
\be \label{e3}(\Delta_\Omega -2)C_{2,1,1} = 9 E_4 - E_2^2\ee
which shall be very useful to us.
 We have that
\be \label{8} I_{D^8 \mathcal{R}^4}^{(1)} = 2\pi\int_{\mathcal{F}_L}  \frac{d^2\Omega}{\Omega_2^2} \Big(C_{2,1,1} + \frac{1}{2} E_2^2 -\frac{1}{2} E_4\Big) \Gamma_{d,d;1}, \ee
on using \C{e1} and \C{e2}. Thus apart from $E_2^2$, the other two terms in the integrand on the right hand side each satisfy an eigenvalue equation on $SL(2,\mathbb{Z})_\Omega$.

Hence on using \C{main}, this leads to
\be \label{int1}\Big(\Delta_{O(d,d,\mathbb{Z})} +\frac{d(d-2)}{2}\Big) \hat{I}_{D^8\mathcal{R}^4}^{(1)} =  4\pi \int_{\mathcal{F}_L}\frac{d^2\Omega}{\Omega_2^2} \Big( C_{2,1,1} - \frac{1}{2}E_4\Big)\Delta_\Omega \Gamma_{d,d;1},\ee
where we have defined
\be \hat{I}_{D^8\mathcal{R}^4}^{(1)} = I_{D^8\mathcal{R}^4}^{(1)} - \pi\int_{\mathcal{F}_L}  \frac{d^2\Omega}{\Omega_2^2} E_2^2  \Gamma_{d,d;1}.\ee
On integrating by parts the terms on the right hand side of \C{int1} and using \C{e3}, we get that
\bea \label{main1}&&\Big(\Delta_{O(d,d,\mathbb{Z})} +\frac{(d+2)(d-4)}{2}\Big) \hat{I}_{D^8\mathcal{R}^4}^{(1)} = 8\pi \int_{\mathcal{F}_L}  \frac{d^2\Omega}{\Omega_2^2} \Big(2E_4 -\frac{1}{2} E_2^2 \Big)\Gamma_{d,d;1} \non \\ &&+4\pi \int_{-1/2}^{1/2} d\Omega_1 \Big[\Big( C_{2,1,1} - \frac{1}{2}E_4\Big) \frac{\p\Gamma_{d,d;1}}{\p\Omega_2}- \Gamma_{d,d;1} \frac{\p}{\p\Omega_2} \Big( C_{2,1,1} - \frac{1}{2}E_4\Big)\Big]\Big\vert_{\Omega_2 = L\rightarrow \infty}.\eea

To calculate the boundary contribution, we use the alternate expression for the lattice factor given by
\be \label{altgenlat}\Gamma_{d,d;1} (G,B;\Omega)= \Omega_2^{d/2} \sum_{m^i, n^i \in \mathbb{Z}} e^{-\pi \Omega_2 \mathcal{L} +2\pi i m^i  n^i \Omega_1},\ee   
where
\be \mathcal{L} = G^{ij} (m^i + B_{ik} n^k)(m^j + B_{jl} n^l) + G_{ij} n^i n^j.\ee
Thus as $\Omega_2 = L \rightarrow \infty$, only the $m^i= n^i =0$ term in the sum contributes.

To obtain these contributions, we use the expressions~\cite{D'Hoker:2015foa}
\bea \label{asymp1}\pi^4 E_4 &=& 2\zeta(8) \Omega_2^4 + \frac{5\pi}{8 \Omega_2^3} \zeta(7) , \non \\ \pi^4 C_{2,1,1} &=& \frac{4}{3}\zeta(8)\Omega_2^4 +2\pi\zeta(3)\zeta(4)\Omega_2 + \frac{5\pi}{2\Omega_2} \zeta(2)\zeta(5) - \frac{3}{2 \Omega_2^2} \zeta(2)\zeta(3)^2 +\frac{9\pi}{16 \Omega_2^3} \zeta(7)\non \\ \eea
where we have ignored terms that vanish exponentially as $\Omega_2 \rightarrow \infty$. Thus for $d\leq 7$, we get a finite contribution
\bea &&4\pi \int_{-1/2}^{1/2} d\Omega_1 \Big[\Big( C_{2,1,1} - \frac{1}{2}E_4\Big) \frac{\p\Gamma_{d,d;1}}{\p\Omega_2}- \Gamma_{d,d;1} \frac{\p}{\p\Omega_2} \Big( C_{2,1,1} - \frac{1}{2}E_4\Big)\Big]\Big\vert_{\Omega_2 = L\rightarrow \infty} \non \\ &&= 5\zeta(5) \delta_{d,4} - \frac{5}{\pi}\zeta(3)^2 \delta_{d,6}.\eea

Next consider the contribution involving $E_4$ in the integrand in \C{main1}. Defining
\be J_1 = \int_{\mathcal{F}_L}  \frac{d^2\Omega}{\Omega_2^2} E_4 \Gamma_{d,d;1} \ee
and using \C{main}, we have that
\bea \Big(\Delta_{O(d,d,\mathbb{Z})} +\frac{(d+6)(d-8)}{2}\Big) J_1  =2\int_{-1/2}^{1/2} d\Omega_1 \Big[E_4 \frac{\p\Gamma_{d,d;1}}{\p\Omega_2}- \Gamma_{d,d;1} \frac{\p E_4}{\p\Omega_2} \Big]\Big\vert_{\Omega_2 = L\rightarrow \infty}.\eea
The right hand side vanishes for $d \leq 7$, leading to
\be \label{defJ1}\Big(\Delta_{O(d,d,\mathbb{Z})} +\frac{(d+6)(d-8)}{2}\Big) J_1  =0.\ee
Hence we see that $J_1$ satisfies Laplace equation as the source terms given by contributions from the boundary of moduli space vanish.

Thus we have that
\bea \label{gend}\Big(\Delta_{O(d,d,\mathbb{Z})} +\frac{(d+2)(d-4)}{2}\Big) \hat{I}_{D^8\mathcal{R}^4}^{(1)} &=& 16\pi J_1 -4\pi\int_{\mathcal{F}_L}  \frac{d^2\Omega}{\Omega_2^2}  E_2^2 \Gamma_{d,d;1} \non \\ &&+5\zeta(5) \delta_{d,4} - \frac{5}{\pi}\zeta(3)^2 \delta_{d,6}. \eea

Unlike the coefficient functions of the various BPS interactions, the relevant integrals are not just boundary contributions from moduli space which lead to considerable complications in general. This shows how the structure of amplitudes not protected by supersymmetry is drastically different from their counterparts that are protected.  

\subsection{The one loop $D^8\mathcal{R}^4$ interaction in nine dimensions}

While \C{gend} can be analyzed in arbitrary dimensions, we shall consider the simplest case of compactifying on a circle. This simple case will be rich enough to illustrate the method and see the complications that arise compared to BPS interactions.    
Thus using
\be \label{one}\Delta_{O(1,1,\mathbb{Z})} =  \frac{1}{2}\Big(r^2\frac{d^2}{d r^2} +r\frac{d}{d r} \Big)\ee
where $r$ is the radius of the circle in the string frame, from \C{gend} we have that
\be  \label{gen2}\frac{1}{2}\Big(r^2\frac{d^2}{d r^2} +r\frac{d}{d r}-9\Big) \hat{I}_{D^8\mathcal{R}^4}^{(1)} = 16\pi J_1 -4\pi\int_{\mathcal{F}_L}  \frac{d^2\Omega}{\Omega_2^2}  E_2^2 \Gamma_{1,1;1}.\ee

We now consider the terms on the right hand side of \C{gen2}. The term involving $J_1$ can be either directly solved using \C{defJ1} and fixing the overall normalization by explicitly calculating the coefficient of the term with a specific $r$ dependence, or it can be evaluated using the unfolding technique. 

We mention the unfolding technique as it will be useful for evaluating other terms that arise in the analysis. This technique gives us that~\cite{McClain:1986id,O'Brien:1987pn,Ditsas:1988pm}
\be \label{rhs}\int_{\mathcal{F}_L}  \frac{d^2\Omega}{\Omega_2^2}  f(\Omega,\bar\Omega) \Gamma_{1,1;1} = r \Big[ \int_{\mathcal{F}_L}  \frac{d^2\Omega}{\Omega_2^2} + \sum_{m\neq 0} \int_{-1/2}^{1/2} d\Omega_1 \int_0^L \frac{d\Omega_2}{\Omega_2^2} e^{-\pi r^2 m^2/\Omega_2}\Big] f(\Omega,\bar\Omega)\ee 
where $f(\Omega,\bar\Omega)$ is modular invariant. 

For $f=E_4$, the first term in \C{rhs} which is linear in $r$ vanishes as it does not leave a finite remainder as $L\rightarrow \infty$. Evaluating the second term which leaves a finite remainder, we get that   
\be J_1 = \frac{15}{2\pi^7}\zeta(7)\zeta(8) \Big( r^7+\frac{1}{r^7}\Big).\ee
We readily make use of
\be \zeta(2s-1) \Gamma(s-1/2) = \pi^{2s-3/2} \zeta(2-2s)\Gamma(1-s)\ee
in our analysis at various places. 

Let us now consider the only remaining term
\be \mathcal{K}_1 = \int_{\mathcal{F}_L}  \frac{d^2\Omega}{\Omega_2^2}  E_2^2 \Gamma_{1,1;1}\ee
in \C{gen2} which we evaluate using \C{rhs}. Note that $E_2^2$ does not satisfy Laplace equation on $SL(2,\mathbb{Z})_\Omega$, hence using \C{main} is not particularly useful for our purposes. The integrands for BPS amplitudes only involve $SL(2,\mathbb{Z})_\Omega$ invariant expressions which satisfy Laplace equation on the fundamental domain of $SL(2,\mathbb{Z})_\Omega$. This is the essential reason for their simplicity. For $\mathcal{K}_1$ to evaluate the first term on the right hand side of \C{rhs}, we note that~\cite{Zagier2,Green:1999pv,Green:2008uj}
\be \label{rhs2}\frac{\pi^{2s}}{4\zeta(2s)^2} \int_{\mathcal{F}_L} \frac{d^2\Omega}{\Omega_2^2} E_s^2 = \frac{L^{2s-1}}{2s-1} +2\phi(s){\rm ln}\Big(\frac{L}{\mu_{2s}}\Big)+\ldots\ee
for $s> 1/2$, and where we have dropped terms that vanish as $L\rightarrow \infty$. In \C{rhs2}, we have that
\bea \phi (s) &=&  \sqrt{\pi} \frac{\Gamma(s-1/2)\zeta(2s-1)}{\Gamma (s)\zeta(2s)},\non \\ {\rm ln}\mu_{2s} &=&\frac{\zeta'(2s-1)}{\zeta(2s-1)} -\frac{\zeta'(2s)}{\zeta(2s)} +\frac{\Gamma'(s-1/2)}{2\Gamma(s-1/2)} -\frac{\Gamma'(s)}{2\Gamma(s)}.\eea
Thus on unfolding, this yields a finite contribution to $\mathcal{K}_1$ as $L\rightarrow \infty$ given by
\be -\frac{4r}{\pi^3} \zeta(3)\zeta(4){\rm ln}\mu_4,\ee
where
\be {\rm ln}\mu_4 = \frac{\zeta'(3)}{\zeta(3)} -\frac{\zeta'(4)}{\zeta(4)} +\frac{1}{2} -{\rm ln} 2,\ee
on using
\be \frac{\Gamma'(3/2)}{\Gamma(3/2)}= 2-\gamma-{\rm ln}4, \quad \frac{\Gamma'(2)}{\Gamma(2)} = 1-\gamma.\ee
The remaining contribution to $\mathcal{K}_1$ from the second term on the right hand side of \C{rhs} is given by
\bea \label{z}\frac{r}{\pi^4}\sum_{m\neq 0}  \int_0^L \frac{d\Omega_2}{\Omega_2^2} e^{-\pi m^2 r^2/\Omega_2}\Big[ 4 \zeta(4)^2 \Omega_2^4 + \frac{\pi^2}{\Omega_2^2} \zeta(3)^2 +4\pi\zeta(3)\zeta(4)\Omega_2 \non \\ + 32 \pi^4\Omega_2  \sum_{k=1}^\infty k^3 \mu(k,2)^2 K_{3/2}^2 (2\pi k\Omega_2)\Big]\eea
on using the expression for the Eisenstein series in \C{Eisenstein} and integrating over $\Omega_1$. 

First consider the contributions from the three terms which are power behaved in $\Omega_2$ in \C{z}. The first two terms give us
\be \frac{15}{\pi^7} \zeta(4)^2\zeta(7) r^7 + \frac{4}{\pi^5 r^5} \zeta(3)^2 \zeta(6) \ee
hence yielding non--vanishing remainders as $L\rightarrow \infty$. On the other hand, the third term has a logarithmic divergence as $L\rightarrow \infty$ and gives us 
\be \label{in}\frac{8r}{\pi^3}\zeta(3)\zeta(4) \sum_{m=1}^\infty \Gamma(0,\pi r^2 m^2/L)\ee
which we consider as $L\rightarrow \infty$. Using the small $z$ expansion 
\be \Gamma (0,z) = -{\rm ln} z -\g +O(z)\ee
for the incomplete Gamma function, we see that \C{in} gives us a finite contribution
\be\frac{4r}{\pi^3} \zeta(3)\zeta(4) \Big( 2{\rm ln}r +\gamma -{\rm ln} (4\pi)\Big)\ee
where we have used
\be \zeta(0) = -1/2, \quad 2\zeta'(0) = - {\ln} (2\pi).\ee
Finally the remaining terms in \C{z} which are exponentially suppressed at large $\Omega_2$ contribute
\be \frac{64}{\pi^3 r^3} \sum_{k,m=1}^\infty \frac{k}{m^4} \mu(k,2)^2 \Big[ \pi rm \sqrt{k}(1+\pi^2 r^2 m^2 k) K_1 (4\pi mr\sqrt{k}) + (1+2\pi^2 r^2 m^2 k) K_2 (4\pi mr\sqrt{k})\Big]\ee
to $\mathcal{K}_1$. Putting together all these contributions, we have the complete expression for $\mathcal{K}_1$ in nine dimensions. 

Using these various expressions, we now solve \C{gen2} to obtain the coefficient function of the $D^8\mathcal{R}^4$ interaction\footnote{Recall that differential equations of the form
\be \label{eE}\frac{d^2 I(r)}{dr^2} + g_1 (r) \frac{dI(r)}{dr} + g_2 (r) I(r) = h(r)\ee
have a particular solution $I_p(r)$ given by
\be I_p(r) = -y_1 (r) \int dr \frac{y_2 h}{W} + y_2 (r)\int dr \frac{y_1 h}{W}\ee
where $y_1 (r)$ and $y_2 (r)$ are the two solutions to the homogeneous equation \C{eE} with $h(r)=0$. Here $W$ is the Wronskian defined by
\be W = y_1 \frac{dy_2}{dr} - y_2 \frac{dy_1}{dr}.\ee
Thus the complete solution is given by
\be  I(r) = c_1 y_1(r) + c_2 y_2(r) + I_p(r),\ee
where $c_1$ and $c_2$ are undetermined coefficients.}. We get that
\bea && I_{D^8\mathcal{R}^4}^{(1)} = c_1 r^3 +\frac{c_2}{r^3} + \frac{20}{\pi^6} \zeta(7)\zeta(8) r^7 + \frac{8}{\pi^2} \zeta(3)\zeta(4)r \Big({\rm ln} (\mu r^2) +\frac{1}{4}\Big)+ \frac{2}{\pi^4 r^5} \zeta(3)^2\zeta(6) \non \\ && +\frac{6}{\pi^6 r^7} \zeta(7)\zeta(8) -\frac{4}{3 \pi^2 r^3} \sum_{k,m=1}^\infty \frac{k}{m^4}  \mu(k,2)^2 \x^6\int \frac{d\x}{\x^6} \Big[ (16+\x^2) K_1 (\x) + \frac{8}{\x} (8+\x^2) K_2 (\x)\Big]\non \\ && +\frac{64}{\pi^2 r^3} \sum_{k,m=1}^\infty \frac{k}{m^4} \mu(k,2)^2\Big[ \frac{1}{3} \Big(1 + \frac{5\x^2}{16}\Big)K_0(\x) +\Big(\frac{\x^3}{64} +\frac{7\x}{24} + \frac{2}{3\x}\Big) K_1 (\x)\Big],\eea
where $c_1$ and $c_2$ are undetermined constants, 
\be \x = 4\pi \sqrt{k}mr,\ee
and
\be {\rm ln}\mu = \frac{\zeta'(4)}{\zeta(4)} - \frac{\zeta'(3)}{\zeta(3)} -\frac{1}{2} +\g -{\rm ln} 2\pi.\ee

The remaining $\x$ integrals can be done leading to
\bea &&\int \frac{d\x}{\x^6} \Big[ (16+\x^2) K_1 (\x) + \frac{8}{\x} (8+\x^2) K_2 (\x)\Big] \non \\ &&= -\frac{K_0(\x)}{12\x^2} +\frac{1}{60}\int \frac{d\x}{\x}K_2(\x) -\frac{13}{720} \int \frac{d\x}{\x}K_4(\x) \non \\ &&+\frac{1}{840} \int \frac{d\x}{\x} K_6(\x) +\frac{1}{5040} \int \frac{d\x}{\x} K_8(\x)\non \\ &&= -\frac{K_0(\x)}{12\x^2} -\frac{75 K_1 (\x) -81 K_3(\x) +5K_5(\x) +K_7(\x)}{2880\x}\non \\ &&=-4(4+\x^2)\frac{K_1(\x)}{\x^7}-(8+\x^2)\frac{K_0(\x)}{\x^6}.\eea
The details relevant to the evaluation are mentioned in the appendix.
Note that
\be \int \frac{d\x}{\x} K_0(\x)\ee
cancels in the final answer, which does not produce a simple expression in terms of Bessel functions.

Finally, this leads to
\bea \label{8final} && I_{D^8\mathcal{R}^4}^{(1)} = c_1 r^3 +\frac{c_2}{r^3} + \frac{20}{\pi^6} \zeta(7)\zeta(8) r^7 + \frac{8}{\pi^2} \zeta(3)\zeta(4)r \Big({\rm ln} (\mu r^2) +\frac{1}{4}\Big)+ \frac{2}{\pi^4 r^5} \zeta(3)^2\zeta(6) \non \\  &&+\frac{6}{\pi^6 r^7} \zeta(7)\zeta(8)  +\frac{4}{\pi^2 r^3} \sum_{k,m=1}^\infty \frac{k}{m^4} \mu(k,2)^2\Big[  2\Big(\x^2 +4\Big)K_0(\x) +\Big(\x^4 +24 \x^2 + 64\Big) \frac{K_1 (\x)}{4\x} \Big].\non \\ \eea

We now fix the coefficients $c_1$ and $c_2$. To do so we directly use \C{rhs} with $f(\Omega,\bar\Omega)$ in \C{8}. As the first term on the right hand side of \C{rhs} is linear in $r$, we ignore it. For the contribution from the second term, after integrating over $\Omega_1$ we consider the various terms in the integrand in the large $\Omega_2$ limit. The terms which are exponentially suppressed in $C_{2,1,1} +E_2^2/2 - E_4/2$ are all of the form
\be \Omega_s^a e^{-2\pi b \Omega_2}.\ee
On integrating over $\Omega_2$, they all yield Bessel functions of the form $K_p (qr)$ on using the integral representation
\be K_s (x) = \frac{1}{2} \Big(\frac{x}{2}\Big)^s \int_0^\infty \frac{dt}{t^{s+1}} e^{-t- x^2/4t}\ee
and hence are exponentially suppressed at large $r$. Hence they can be neglected as well. Thus the terms we need can only come from terms that are power law behaved in $\Omega_2$ in $C_{2,1,1} +E_2^2/2 - E_4/2$. In fact, simple scaling shows that $r^3$ and $r^{-3}$ terms in the final expression can only come $\Omega_2^2$ and $\Omega_2^{-1}$ terms in $C_{2,1,1} +E_2^2/2 - E_4/2$. However, from \C{asymp1} and the expression for $E_2$ we see there are no $\Omega_2^2$  terms and hence
\be c_1 =0.\ee 
Also the $\Omega_2^{-1}$ term in $C_{2,1,1}$ is non--vanishing and we get that
\be c_2 = \frac{\zeta(2)\zeta(5)}{9}.\ee 

\subsection{Possible implications for U--duality from the harmonic anomaly}

Let us consider the presence of the last two terms in \C{gend} which are non--vanishing only in specific dimensions. These harmonic anomalies suggest the presence of some non--analytic terms in the effective action based on analogy with the other known BPS amplitudes and the general structure of the effective action. 

Let $\hat{\mathcal{E}}_{D^{2k}\mathcal{R}^4}$ be the coefficient of the U--duality invariant $D^{2k}\mathcal{R}^4$ term in the Einstein frame. We denote by $\mathcal{E}_{D^{2k}\mathcal{R}^4}$ the perturbative contributions it receives. Thus $\mathcal{E}_{D^{2k}\mathcal{R}^4}$  depends only on $G_{ij}$, $B_{ij}$ and the T--duality invariant dilaton. For compactification on $T^d$, we have that
\be \label{def1}  \mathcal{E}_{D^8\mathcal{R}^4} (G,B,g_d) = \mathcal{E}_{D^8\mathcal{R}^4}^{an}(G,B,g_d)+ \mathcal{E}_{D^8\mathcal{R}^4}^{non-an}(G,B,g_d) ,\ee
where $g_d^{-2} = e^{-2\phi_d} = e^{-2\phi} V_d$ is the T--duality invariant string coupling. The two terms on the right hand side of \C{def1} are the analytic and non--analytic contributions in the string coupling $g_d$ respectively. 
In particular,
\bea \label{8exp}\mathcal{E}^{an}_{D^8\mathcal{R}^4} (G,B,g_d)= g_d^{28/(d-8)} I^{(0)}_{D^8\mathcal{R}^4}(G,B) + g_d^{(2d+12)/(d-8)} I^{(1)}_{D^8\mathcal{R}^4} (G,B) +\ldots. \eea
Now the structure of \C{gend} suggests that among other contributions we should have that 
\be \label{nl}\mathcal{E}_{D^8\mathcal{R}^4}^{non-an} \sim (\delta_{d,4}\mathcal{E}_{D^4\mathcal{R}^4}  +\delta_{d,6} \mathcal{E}_{D^6\mathcal{R}^4}) {\rm ln} g_d.\ee  
This is because the U-duality invariant equation must be of the form\footnote{The coefficient function can split into a sum of coefficient functions each of which satisfies a U--duality invariant equation. Our analysis does not depend on this detail.} 
\be \Delta_U \hat{\mathcal{E}}_{D^8\mathcal{R}^4} = \ldots,\ee 
where $\Delta_U$ is the U--duality invariant Laplacian. On using
\be \label{defL}\Delta_U = \frac{8-d}{8} \p_{\phi_d}^2 + \frac{d^2-d+4}{4} \p_{\phi_d} +\Delta_{O(d,d;\mathbb{Z})}+\ldots\ee 
where we have dropped terms involving the R--R moduli, and using
\bea \label{def2}\mathcal{E}_{D^4\mathcal{R}^4}= g_d^{20/(d-8)} I^{(0)}_{D^4\mathcal{R}^4}+g_d^{(2d+4)/(d-8)}I^{(1)}_{D^4\mathcal{R}^4}+\ldots, \non \\ \mathcal{E}_{D^6\mathcal{R}^4} =g_d^{24/(d-8)} I^{(0)}_{D^6\mathcal{R}^4}+g_d^{(2d+8)/(d-8)}I^{(1)}_{D^6\mathcal{R}^4}+\ldots,\eea
we get that
\be \Delta_{O(d,d;\mathbb{Z})} I^{(1)}_{D^8\mathcal{R}^4} \sim \delta_{d,4}I^{(0)}_{D^4\mathcal{R}^4}  +\delta_{d,6} I^{(0)}_{D^6\mathcal{R}^4} \sim \zeta(5) \delta_{d,4} +\zeta(3)^2 \delta_{d,6}\ee
which is precisely the structure of the harmonic anomaly in \C{gend}. These contributions in \C{nl} involving terms logarithmic in the string coupling in the Einstein frame arise from non--analytic terms logarithmic in the external momenta in the string frame on converting from the string frame to the Einstein frame. Thus it is natural to assume that \C{nl} leads to a contribution 
\be \label{nl2}\hat{\mathcal{E}}_{D^8\mathcal{R}^4} \sim (\delta_{d,4}\hat{\mathcal{E}}_{D^4\mathcal{R}^4}  +\delta_{d,6} \hat{\mathcal{E}}_{D^6\mathcal{R}^4}) {\rm ln} g_d\ee
to the complete $D^8\mathcal{R}^4$ interaction. In particular, among other terms \C{nl2} yields~\cite{D'Hoker:2014gfa,Pioline:2015yea} 
\bea \label{nl3}\delta_{d,4}  \Big(\mathcal{E}^{an}_{D^4\mathcal{R}^4} +\mathcal{E}^{an}_{\mathcal{R}^4} {\rm ln}g_d \Big) {\rm ln }g_d +\delta_{d,6}\Big(\mathcal{E}^{an}_{D^6\mathcal{R}^4} + \frac{5}{\pi} \mathcal{E}^{an}_{D^4\mathcal{R}^4}\Big){\rm ln} g_d.\eea 
From \C{nl3} it follows that in six dimensions the $D^8\mathcal{R}^4$ interaction in the string frame receives non--local contributions schematically of the form\footnote{Note that
\be \mathcal{E}_{\mathcal{R}^4} = g_d^{12/(d-8)} I^{(0)}_{\mathcal{R}^4} +g_d^{(2d-4)/(d-8)}I^{(1)}_{\mathcal{R}^4}+\ldots.\ee}
\be {\rm ln}(\alpha' s) \Big(I^{(0)}_{D^4\mathcal{R}^4} + g_4^2 I^{(1)}_{D^4\mathcal{R}^4} + g_4^4 I^{(2)}_{D^4\mathcal{R}^4}\Big) + g_4^2{\rm ln}^2 (\alpha' s) \Big( I^{(0)}_{\mathcal{R}^4} +g_4^2 I^{(1)}_{\mathcal{R}^4}\Big)\ee
at one, two and three loops, coming from the integral over $\mathcal{R}_L$. Here $s$ is a generic Mandelstam variable.
Similarly it receives non--local contributions in four dimensions in the string frame of the form
\be  {\rm ln}(\alpha' s) \Big(I^{(0)}_{D^6\mathcal{R}^4} + g_6^2 I^{(1)}_{D^6\mathcal{R}^4} + g_6^4 I^{(2)}_{D^6\mathcal{R}^4} + g_6^6 I^{(3)}_{D^6\mathcal{R}^4}\Big) + g_6^2{\rm ln}^2 (\alpha' s) \Big( I^{(0)}_{D^4 \mathcal{R}^4} +g_6^2 I^{(1)}_{D^4\mathcal{R}^4}+ g_6^4 I^{(2)}_{D^4\mathcal{R}^4}\Big).\ee 
at one, two, three and four loops. It would be interesting to see if these expectations are indeed borne out by explicit calculations. 

\section{The one loop $D^{10}\mathcal{R}^4$ interaction}

We next consider the $D^{10} \mathcal{R}^4$ interaction that arises from the low energy expansion of the one loop four graviton amplitude. Since most of the arguments are similar to those for the analysis of the $D^8\mathcal{R}^4$ interaction,we shall be somewhat brief.  
Among the various terms in $j^{(1,1)}$ in \C{list}, $D_2$, $D_3$, $D_{1,1,1}$ and $D_{1,1,1,1;1}$ are simple and are given by \C{e1} and~\cite{D'Hoker:2015foa} 
\be \label{e4}D_{1,1,1} = E_3, \quad D_3 = E_3 +\zeta(3), \quad D_{1,1,1,1;1} = \frac{2}{5} E_5 +\frac{\zeta(5)}{30}.\ee
However, $D_{3,1,1}$, $D_5$ and $D_{2,2,1}$ are more involved and are given by\footnote{These equations and the Poisson equation \C{e6} have been conjectured in~\cite{D'Hoker:2015foa}.}
\bea \label{e5}
40 D_{3,1,1} &=& 300 D_{2,1,1,1} + 120 E_2 E_3 -276 E_5 +7\zeta(5), \non \\ 
D_5 &=& 60 D_{2,1,1,1} +10 E_2 E_3 - 48 E_5 +10\zeta(3) E_2 + 16\zeta(5), \non \\ 
10 D_{2,2,1} &=& 20 D_{2,1,1,1} - 4 E_5 + 3\zeta(5),\eea
where $D_{2,1,1,1}$ satisfies the Poisson equation
\be \label{e6}(\Delta_\Omega -6) D_{2,1,1,1} =  \frac{86}{5} E_5 - 4E_2 E_3 +\frac{\zeta(5)}{10}.\ee
Thus for the $D^{10}\mathcal{R}^4$ interaction we have that
\be \label{10} I_{D^{10} \mathcal{R}^4}^{(1)} = \frac{10\pi}{6!}\int_{\mathcal{F}_L}  \frac{d^2\Omega}{\Omega_2^2} \Big(336 D_{2,1,1,1} + 240 E_2 E_3 -\frac{1632}{5}E_5 +48 \zeta(3) E_2 +\frac{144}{5} \zeta(5)\Big) \Gamma_{d,d;1}. \ee

On using \C{main} we have that
\bea &&\Big(\Delta_{O(d,d,\mathbb{Z})} +\frac{d(d-2)}{2}\Big) \hat{I}_{D^{10}\mathcal{R}^4}^{(1)} \non \\
&&=  \frac{20\pi}{6!}\int_{\mathcal{F}_L}  \frac{d^2\Omega}{\Omega_2^2} \Big(336 D_{2,1,1,1} -\frac{1632}{5}E_5 +48 \zeta(3) E_2 +\frac{144}{5} \zeta(5)\Big)\Delta_\Omega\Gamma_{d,d;1},\eea
where
\be  \hat{I}_{D^{10}\mathcal{R}^4}^{(1)} =  I_{D^{10}\mathcal{R}^4}^{(1)}- \frac{10\pi}{3}\int_{\mathcal{F}_L}  \frac{d^2\Omega}{\Omega_2^2}E_2 E_3 \Gamma_{d,d;1}. \ee
Integrating by parts and using \C{e6}, we obtain the differential equation
\bea \label{MaiN}&&\Big(\Delta_{O(d,d,\mathbb{Z})} +\frac{(d+4)(d-6)}{2}\Big) \hat{I}_{D^{10}\mathcal{R}^4}^{(1)} \non \\ &&=\pi \int_{\mathcal{F}_L}  \frac{d^2\Omega}{\Omega_2^2} \Big(\frac{168}{5}E_5 -\frac{112}{3} E_2 E_3 -\frac{16}{3} \zeta(3)E_2 - \frac{58}{15}\zeta(5) \Big)\Gamma_{d,d;1}\non \\ &&+\pi \int_{-1/2}^{1/2} d\Omega_1 \Big[\Big( \frac{28}{3} D_{2,1,1,1} - \frac{136}{15} E_5 +\frac{4}{3} \zeta(3) E_2 +\frac{4}{5} \zeta(5)\Big) \frac{\p\Gamma_{d,d;1}}{\p\Omega_2}\non \\ &&- \Gamma_{d,d;1} \frac{\p}{\p\Omega_2} \Big( \frac{28}{3} D_{2,1,1,1} - \frac{136}{15} E_5 +\frac{4}{3} \zeta(3) E_2 \Big)\Big]\Big\vert_{\Omega_2 = L\rightarrow \infty}.\eea

To calculate the boundary contribution for $d \leq 7$, we again use \C{altgenlat} and keep finite terms as $\Omega_2 = L\rightarrow \infty$. We use the expressions~\cite{D'Hoker:2015foa}
\bea \label{asymp2}\pi^2 E_2 &=& 2\zeta(4)\Omega_2^2 + \frac{\pi}{\Omega_2}\zeta(3), \non \\ \pi^5 E_5 &=& 2\zeta(10) \Omega_2^5 + \frac{35\pi}{64\Omega_2^4} \zeta(9) , \non \\ \pi^5 D_{2,1,1,1} &=& \frac{6}{5}\zeta(10) \Omega_2^5 + 2\pi \zeta(3)\zeta(6) \Omega_2^2 - \frac{\pi}{2}\zeta(4)\zeta(5) + \frac{21\pi}{8\Omega_2^2}  \zeta(2)\zeta(7) \non \\ &&- \frac{3}{\Omega_2^3} \zeta(2)\zeta(3)\zeta(5)  +\frac{43\pi}{64\Omega_2^4} \zeta(9)
\eea
where we have dropped terms that are exponentially suppressed at large $\Omega_2$.
 
Thus for $d \leq 7$ we get a finite contribution to \C{MaiN} given by
\bea &&\pi \int_{-1/2}^{1/2} d\Omega_1 \Big[\Big( \frac{28}{3} D_{2,1,1,1} - \frac{136}{15} E_5 +\frac{4}{3} \zeta(3) E_2 +\frac{4}{5} \zeta(5)\Big) \frac{\p\Gamma_{d,d;1}}{\p\Omega_2}\non \\ &&- \Gamma_{d,d;1} \frac{\p}{\p\Omega_2} \Big( \frac{28}{3} D_{2,1,1,1} - \frac{136}{15} E_5 +\frac{4}{3} \zeta(3) E_2 \Big)\Big]\Big\vert_{\Omega_2 = L\rightarrow \infty} \non \\ &&= \frac{101\pi}{135}\zeta(5) \delta_{d,2} +4\zeta(3)^2 \delta_{d,4} +\frac{245}{12\pi}\zeta(7)\delta_{d,6}.\eea

Next consider the contribution involving $E_5$ in the integrand in \C{MaiN}. Defining
\be J_2 = \int_{\mathcal{F}_L}  \frac{d^2\Omega}{\Omega_2^2} E_5 \Gamma_{d,d;1}, \ee
we have that
\bea \Big(\Delta_{O(d,d,\mathbb{Z})} +\frac{(d+8)(d-10)}{2}\Big) J_2  =2\int_{-1/2}^{1/2} d\Omega_1 \Big[E_5 \frac{\p\Gamma_{d,d;1}}{\p\Omega_2}- \Gamma_{d,d;1} \frac{\p E_5}{\p\Omega_2} \Big]\Big\vert_{\Omega_2 = L\rightarrow \infty}.\eea
The right hand side vanishes for $d \leq 7$, leading to
\be \label{Gen1}\Big(\Delta_{O(d,d,\mathbb{Z})} +\frac{(d+8)(d-10)}{2}\Big) J_2  =0.\ee
Thus we obtain the Poisson equation
\bea \label{gend2}&&\Big(\Delta_{O(d,d,\mathbb{Z})} +\frac{(d+4)(d-6)}{2}\Big) \hat{I}_{D^{10}\mathcal{R}^4}^{(1)} =\frac{168\pi}{5}J_2 -\frac{112\pi}{3} \int_{\mathcal{F}_L}  \frac{d^2\Omega}{\Omega_2^2} E_2 E_3 \Gamma_{d,d;1} \non \\ &&-\frac{8}{3} \zeta(3)I_{D^4\mathcal{R}^4}^{(1)} - \frac{29}{15}\zeta(5)I_{\mathcal{R}^4}^{(1)} +\frac{101\pi}{135}\zeta(5) \delta_{d,2} +4\zeta(3)^2 \delta_{d,4} +\frac{245}{12\pi}\zeta(7)\delta_{d,6}\eea
on using \C{1genus1} and \C{1genus2}.

\subsection{The one loop $D^{10}\mathcal{R}^4$ interaction in nine dimensions}

We analyze \C{gend2} in nine dimensions which is the simplest case. Using \C{one}, in nine dimensions we have that
\bea  \label{9}\frac{1}{2}\Big(r^2\frac{d^2}{d r^2} +r\frac{d}{d r}-25\Big) \hat{I}_{D^{10}\mathcal{R}^4}^{(1)} &=& \frac{168\pi}{5}J_2 -\frac{112\pi}{3} \int_{\mathcal{F}_L}  \frac{d^2\Omega}{\Omega_2^2} E_2 E_3 \Gamma_{1,1;1} \non \\ &&-\frac{8}{3} \zeta(3)I_{D^4\mathcal{R}^4}^{(1)} - \frac{29}{15}\zeta(5)I_{\mathcal{R}^4}^{(1)}.\eea
Proceeding as in the case of the $D^8\mathcal{R}^4$ interaction, we get that
\be J_2 = \frac{105}{4\pi^9}\zeta(9)\zeta(10) \Big( r^9+\frac{1}{r^9}\Big).\ee
Also we obtain the expressions for the $\mathcal{R}^4$ and $D^4\mathcal{R}^4$ interactions from \C{9d}.

Let us now consider the remaining expression
\be \mathcal{K}_2 = \int_{\mathcal{F}_L}  \frac{d^2\Omega}{\Omega_2^2}  E_2 E_3 \Gamma_{1,1;1}\ee
in \C{9} which we evaluate using the unfolding technique in \C{rhs}.
The first term on the right hand side of \C{rhs} does not give any contribution as $L\rightarrow \infty$.
The remaining contribution to $\mathcal{K}_2$ is given by
\bea \label{L} \frac{r}{\pi^5}\sum_{m\neq 0}  \int_0^L \frac{d\Omega_2}{\Omega_2^2} e^{-\pi m^2 r^2/\Omega_2}\Big[ 4\zeta(4)\zeta(6)\Omega_2^5 + 2\pi\zeta(3)\zeta(6) \Omega_2^2 + \frac{3\pi}{2} \zeta(4)\zeta(5) \non \\ + \frac{3\pi^2}{4 \Omega_2^3} \zeta(3)\zeta(5)    + 16 \pi^5\Omega_2  \sum_{k=1}^\infty k^4 \mu(k,2) \mu(k,3) K_{3/2} (2\pi k\Omega_2) K_{5/2} (2\pi k\Omega_2)\Big]\eea
on using the expression for the Eisenstein series in \C{Eisenstein} and integrating over $\Omega_1$. 
The first four terms which are power behaved in $\Omega_2$ give us the finite contributions
\be \frac{105}{2\pi^9} \zeta(4)\zeta(6)\zeta(9) r^9 + \frac{2}{\pi^5} \zeta(3)^2 \zeta(6) r^3 +  \frac{3}{\pi^5 r} \zeta(2)\zeta(4)\zeta(5) +\frac{9}{\pi^7 r^7}\zeta(3)\zeta(5) \zeta(8).\ee
Finally the terms which are exponentially suppressed for large $\Omega_2$ in \C{L} give us
\bea \frac{16}{\pi^5 r^5} \sum_{k,m=1}^\infty \frac{k}{m^6} \mu(k,2) \mu(k,3) \Big[ \pi rm \sqrt{k}\Big(9+12 \pi^2 r^2 m^2 k + 2\pi^4 m^4 r^4 k^2\Big) K_1 (4\pi mr\sqrt{k}) \non \\ + \Big(3+2\pi^2 r^2 m^2 k\Big) \Big(3+4\pi^2 r^2 m^2 k \Big)K_2 (4\pi mr\sqrt{k})\Big].\eea

Including the various contributions, we solve \C{9} to obtain
\bea I_{D^{10}\mathcal{R}^4}^{(1)} &=& d_1 r^5 + \frac{d_2}{r^5} + \frac{147}{\pi^8} \zeta(9)\zeta(10) r^9 +\frac{20}{7\pi^2}\zeta(3)^2\zeta(4) r^3 + \frac{29}{45} \zeta(2)\zeta(5)r \non \\ &&+ \frac{116}{135 r} \zeta(2)\zeta(5) +\frac{4}{3\pi^2 r^3} \zeta(3)^2 \zeta(4) +\frac{2}{\pi^6 r^7} \zeta(3)\zeta(5)\zeta(8) + \frac{63}{2\pi^8 r^9} \zeta(9)\zeta(10) \non \\ &&+\frac{1}{240 \pi^4 r^5} \sum_{k,m=1}^\infty \frac{k}{m^6} \mu(k,2)\mu(k,3)\Big[ 8 \Big(-1728 + 624\x^2 + 43 \x^4 \Big)K_0(\x) \non \\ &&+\Big(-27648+6528 \x^2 + 2080 \x^4 + 25 \x^6 \Big)\frac{K_1(\x)}{\x}\Big] \non \\ &&- \frac{7\cdot 2^8}{15 \pi^4 r^5} \sum_{k,m=1}^\infty \frac{k}{m^6} \mu(k,2)\mu(k,3) \x^{10} \int \frac{d\x}{\x^{10}} \Big[ \Big(\frac{9}{4}+\frac{12\x^2}{4^3} + \frac{2\x^4}{4^5}\Big)K_1 (\x) \non \\ &&+ \Big( 9+\frac{18\x^2}{4^2} +\frac{8\x^4}{4^4}\Big)\frac{K_2(\x)}{\x}\Big]\eea
where $d_1$ and $d_2$ are undetermined constants. This further simplifies using the result
\bea &&\int \frac{d\x}{\x^{10}} \Big[ \Big(\frac{9}{4}+\frac{12\x^2}{4^3} + \frac{2\x^4}{4^5}\Big)K_1 (\x) + \Big( 9+\frac{18\x^2}{4^2} +\frac{8\x^4}{4^4}\Big)\frac{K_2(\x)}{\x}\Big]\non \\ &&= -\Big(3 + \frac{3\x^2}{4}  +\frac{5\x^4}{128} \Big)\frac{K_1(\x)}{2\x^{11}} -\Big(3 +\frac{3\x^2}{8}  +\frac{\x^4}{128}\Big)\frac{K_0(\x)}{4\x^{10}}.\eea

Thus we get that
\bea \label{10final}I_{D^{10}\mathcal{R}^4}^{(1)} &=& d_1 r^5 + \frac{d_2}{r^5} + \frac{147}{\pi^8} \zeta(9)\zeta(10) r^9 +\frac{20}{7\pi^2}\zeta(3)^2\zeta(4) r^3 + \frac{29}{45} \zeta(2)\zeta(5)r \non \\ &&+ \frac{116}{135 r} \zeta(2)\zeta(5) +\frac{4}{3\pi^2 r^3} \zeta(3)^2 \zeta(4) +\frac{2}{\pi^6 r^7} \zeta(3)\zeta(5)\zeta(8) + \frac{63}{2\pi^8 r^9} \zeta(9)\zeta(10) \non \\ &&+\frac{1}{\pi^4 r^5} \sum_{k,m=1}^\infty \frac{k}{m^6} \mu(k,2)\mu(k,3)\Big[  \Big(32 + 32\x^2 + \frac{5 \x^4}{3} \Big)K_0(\x) \non \\ &&+\Big(64+72 \x^2 + 11 \x^4 + \frac{5 \x^6}{48} \Big)\frac{K_1(\x)}{\x}\Big].\eea

We now fix the coefficients $d_1$ and $d_2$, by using \C{rhs} and \C{10}. Arguing as before, the terms we need can only come from terms that are power behaved in $\Omega_2$ in the large $\Omega_2$ expansion in $336 D_{2,1,1,1} +240 E_2 E_3 -1632 E_5/5 +48\zeta(3) E_2$. The $r^5$ and $r^{-5}$ terms in the amplitude can only come $\Omega_2^3$ and $\Omega_2^{-2}$ terms in $336 D_{2,1,1,1} +240 E_2 E_3 -1632 E_5/5 +48\zeta(3) E_2$. However, from \C{asymp2} and the expression for $E_2$ we see there are no $\Omega_2^3$  terms and hence
\be d_1 =0.\ee 
Now the $\Omega_2^{-2}$ term in $D_{2,1,1,1}$ is non--vanishing and we get that
\be d_2 = \frac{7\zeta(2)\zeta(7)}{135}.\ee 

\subsection{Possible implications for U--duality from the harmonic anomaly}

As in the analysis of the $D^8\mathcal{R}^4$ interaction, the presence of the last three terms in \C{gend2} suggest the presence of some non--analytic terms in the string amplitude. Using
\be \mathcal{E}^{an}_{D^{10}\mathcal{R}^4} (G,B,g_d)= g_d^{32/(d-8)} I^{(0)}_{D^{10}\mathcal{R}^4} (G,B)+ g_d^{(2d+16)/(d-8)} I^{(1)}_{D^{10}\mathcal{R}^4} (G,B)+\ldots ,\ee
the structure of \C{gend2} suggests that among other contributions we have that 
\be \hat{\mathcal{E}}_{D^{10}\mathcal{R}^4}^{non-an} \sim (\delta_{d,2}\hat{\mathcal{E}}_{D^4\mathcal{R}^4}  +\delta_{d,4} \hat{\mathcal{E}}_{D^6\mathcal{R}^4} +\delta_{d,6} \hat{\mathcal{E}}_{D^8\mathcal{R}^4}) {\rm ln} g_d.\ee  
This is because the U-duality invariant equation of the form 
\be \label{Ud}\Delta_U \hat{\mathcal{E}}_{D^{10}\mathcal{R}^4} = \ldots,\ee 
on using \C{8exp}, \C{def2} and \C{defL} gives
\be \Delta_{O(d,d;\mathbb{Z})} I^{(1)}_{D^{10}\mathcal{R}^4} \sim \delta_{d,2}I^{(0)}_{D^4\mathcal{R}^4}  +\delta_{d,4} I^{(0)}_{D^6\mathcal{R}^4} +\delta_{d,6} I^{(0)}_{D^8\mathcal{R}^4}\sim \zeta(5) \delta_{d,2} +\zeta(3)^2 \delta_{d,4} +\zeta(7) \delta_{d,6}.\ee
This leads to several consequences for the perturbative contributions to the $D^{10}\mathcal{R}^4$ interaction at various loops along the lines of the previous discussion.  

In fact the source terms involving $\zeta(3) I^{(1)}_{D^4\mathcal{R}^4}$ and $\zeta(5) I^{(1)}_{\mathcal{R}^4}$ in \C{gend2} suggest the presence of source terms $\hat{\mathcal{E}}_{\mathcal{R}^4} \hat{\mathcal{E}}_{D^4\mathcal{R}^4}$ on the right hand side of \C{Ud}.

Thus we have analyzed the contribution to the $D^8\mathcal{R}^4$ and $D^{10} \mathcal{R}^4$ interactions from the low momentum expansion of the four graviton amplitude. We have obtained second order differential equations satisfied by the coefficient functions of these interactions, which we have solved in nine dimensions. Even in this simple setting, note that while the BPS interactions are given by the simple expressions in \C{9d}, the non--BPS ones are complicated and are given by \C{8final} and \C{10final}, elucidating the difference between the interactions protected or not by supersymmetry. These interactions obtained from the one loop four graviton amplitude must be the same in the type IIA and IIB theories. In nine dimensions, this implies invariance under $r \leftrightarrow 1/r$ while keeping $e^{-2\phi}r$ fixed. While this is manifest in \C{9d}, this is not the case in either \C{8final} or \C{10final}. One can extend the analysis to higher orders in the momentum expansion of the one loop four graviton amplitude, on obtaining the detailed structure of the integrands. For the $D^{12}\mathcal{R}^4$ interaction, a modular invariant differential equation for the Mercedes diagram has been obtained~\cite{Basu:2015ayg}, which leads to partial contribution for the $D^{12}\mathcal{R}^4$ interaction. 

It would be interesting to extend the analysis to lower dimensions in detail, and also at higher string loops. This could give insight into the structure of non--BPS operators in maximally supersymmetric string theories, which are not well understood apart from some constraints based on supersymmetry and the explicit multi--loop structure of maximal supergravity~\cite{Green:2008bf,Bern:2008pv,Bern:2009kd,Basu:2013goa,Basu:2014uba,Basu:2015dsa}.      

\appendix

\section{A list of useful formulae}

In the main text, we often need various formulae involving Bessel functions which we summarize. Recall that
\be K_s (\x) = K_{-s} (\x).\ee
We make use of the recurrence identities
\bea K_s (\x) &=& K_{s+2} (\x) -\frac{2(s+1)}{\x} K_{s+1} (\x), \non \\ K_s (\x) &=& K_{s-2} (\x) + \frac{2(s-1)}{\x} K_{s-1} (\x), \non \\ K_s (\x) &=& \frac{\x}{2s} \Big(K_{s+1} (\x)- K_{s-1}(\x)\Big) ,\eea
as well as the integral
\be \int d\x \x^{s+1} K_s (\x) = -\x^{s+1} K_{s+1} (\x).\ee
Some of the simple integrals that are needed for the $D^8\mathcal{R}^4$ interaction are
\bea &&\int d\x K_1 (\x) = -K_0(\x), \quad \int \frac{d\x}{\x} K_2 (\x) = -\frac{K_1(\x)}{\x}, \non \\ && \int d\x \x K_2(\x) = -2 K_0(\x) - \x K_1(\x), \quad \int d\x \x^2 K_1(\x) = -\x^2 K_2 (\x). \eea
For the more involved integrals we use the relation
\be \int \frac{d\x}{\x^m} K_s (\x) = -\frac{K_s(\x)}{(m-1) \x^{m-1}} -\frac{1}{2(m-1)} \int \frac{d\x}{\x^{m-1}} \Big(K_{s-1}(\x) + K_{s+1}(\x)\Big)\ee
where $m$ is an integer greater than 1. Using this iteratively, we obtain integrals with only $\x$ in the denominator and Bessel functions in the numerator. These can be performed using
\be \int \frac{d\x}{\x} K_{2n}(\x) = -\frac{1}{2n} \Big[ (-1)^n K_0(\x) + K_{2n} (\x) +2\sum_{k=1}^{n-1} (-1)^{k+n} K_{2k} (\x)\Big]\ee
where $n$ is a positive integer. Thus we have that
\bea \int \frac{d\x}{\x} K_4 (\x) &=& \frac{1}{2\x} [K_1 (\x) - 3K_3(\x)], \non \\ \int \frac{d\x}{\x} K_6 (\x) &=& -\frac{1}{3\x} [K_1 (\x) - 3K_3(\x) +5K_5(\x)], \non \\ \int \frac{d\x}{\x} K_8 (\x) &=& \frac{1}{4\x} [K_1 (\x) - 3K_3(\x) + 5K_5(\x) - 7K_7 (\x)].\eea

For the $D^{10}\mathcal{R}^4$ amplitude, we also need the simple integrals
\bea \int d\x \x^3 K_2 (\x) &=& -\x^3 K_3 (\x), \non\\ \int d\x \x^4 K_1 (\x) &=& -\x^4 K_4 (\x) +4 \x^3 K_3 (\x).\eea

%\bibliographystyle{utphys}
%\bibliography{myrefs}

\providecommand{\href}[2]{#2}\begingroup\raggedright\endgroup

\end{document}